# PROBING FOR EVIDENCE OF PLUMES ON EUROPA WITH HST/STIS

Sparks, W.B.[1], Hand, K.P.[2], McGrath, M.A.[3], Bergeron, E.[1], Cracraft, M.[1], Deustua, S.E.[1]

[1] Space Telescope Science Institute, 3700 San Martin Drive, Baltimore, MD 21218, USA.

[2] Jet Propulsion Laboratory, 4800 Oak Grove Drive, Pasadena, CA 91109

[3] SETI Institute, 189 N Bernardo Ave, Mountain View, CA 94043

ABSTRACT

Roth et al. (2014a) reported evidence for plumes of water venting from a southern high latitude region on Europa – spectroscopic detection of off-limb line emission from the dissociation products of water. Here, we present *Hubble Space Telescope* (HST) direct images of Europa in the far ultraviolet (FUV) as it transited the smooth face of Jupiter, in order to measure absorption from gas or aerosols beyond the Europa limb. Out of ten observations we found three in which plume activity could be implicated. Two show statistically significant features at latitudes similar to Roth et al., and the third, at a more equatorial location. We consider potential systematic effects that might influence the statistical analysis and create artifacts, and are unable to find any that can definitively explain the features, although there are reasons to be cautious. If the apparent absorption features are real, the magnitude of implied outgassing is similar to that of the Roth et al. feature, however the apparent activity appears more frequently in our data.

KEYWORDS: planets and satellites: general

1. INTRODUCTION

Europa is one of the most compelling astrobiological targets in the Solar System. The combination of geological, compositional, gravity, and induced magnetic field measurements all indicate a contemporary global saline liquid water ocean of ~100 km in depth (Anderson et al., 1998; Pappalardo et al., 1998; Kivelson et al., 2000; Zimmer et al., 2000; Hand & Chyba, 2007). Cycling of the ocean with a silicate seafloor and oxidant laden surface ice could yield an ocean rich with the elements and chemical energy needed to sustain life (Chyba 2000; Chyba & Hand, 2001; Hand et al., 2007; Hand et al., 2009). Europa is therefore one of the most plausible sites in the Solar System beyond Earth where life could exist. With Roth et al. (2014a)'s evidence that plumes of water may be emanating from Europa's surface, the possibility arises that material from Europa's ocean could be ejected out into space and onto Europa's surface, providing direct access to ocean material without the need to drill through the many kilometers of Europa's ice shell. Furthermore, plume activity would indicate that Europa is geologically active in the modern epoch. Therefore, it is of critical importance that additional observations and techniques for assessing plume activity be undertaken.

Europa possesses a tenuous atmosphere and, because of its location deep within the Jovian magnetosphere, auroral activity in the form of far ultraviolet hydrogen and oxygen line emission occurs from the interaction of plasma electrons with the thin gases around the satellite (Hall et al. 1995; McGrath et al. 2009). Though the exosphere of Europa arises from material escaping from the surface, its basic properties such as areal coverage, density, thermal structure, and spatial and temporal variability are poorly understood (McGrath et al., 2009; Roth et al., 2015).

Evidence for off-limb far ultraviolet oxygen and hydrogen emission, dissociation products of water, was inferred by Roth et al. (2014a) to be the result of plume activity. The detection comprises a single 4-$\sigma$ event, and remains to be validated by an independent observational approach (or even the same approach; Roth et al., 2014b). The intriguing possibility that plumes and active cryovolcanism are present on Europa was considered previously (Fagents et al., 2000; Fagents et al., 2003). Fagents et al. (2000) modeled volatile gas and particle-rich plumes in an effort to explain a variety of low-albedo features on Europa. In that work, which also considered the dynamics of Io's observed plumes, the greatest height to which Europa plumes were expected to reach was ~100 km. This height corresponded to gas-dominated plumes with eruption velocities of ~600 m s$^{-1}$. They argue that more realistic values for ejection velocities and plume compositions yield plumes with heights of 1 to < 25 km.

Significantly, the observations by Roth et al. (2014a) necessitate plumes of water vapor that reach heights of ~200 km. This would require an initial eruption velocity of 700 m s$^{-1}$, which is hard to reconcile with the Fagents et al. (2000) work, especially given that the species reaching the highest altitudes in those models were the dissolved and devolatilized gases possibly trapped in the water and ice fractures (e.g. $CO_2$, CO, $NH_3$, $SO_2$). At Enceladus, where the Cassini spacecraft has directly observed plumes (Porco et al., 2006), the water gas velocities are in the range of 300-500 m s$^{-1}$ (Waite et al., 2006; Tian et al., 2007; Brilliantov et al., 2008). For water molecules to achieve velocities of ~700 m s$^{-1}$ on Europa requires vent fractures with surface temperatures of >230 K (Roth et al., 2014a). Though not impossible, such temperature and corresponding plume heights are well above Europa's ~100 K surface temperature, and well below the eutectic temperature for any salt-rich solution that might suppress the freezing temperature of the plume source material (Kargel et al., 2000).

Here we seek to further investigate the existence of plumes on Europa by using a different HST observing strategy than that used by Roth et al; we use HST direct imaging observations of Europa transiting in front of Jupiter, with Europa limb absorptions of Jupiter's reflected light as a possible indication of plumes. In the far ultraviolet, the face of Jupiter is smooth, dominated in appearance by scattering from high level atmospheric hazes in the Jovian atmosphere. Not only that, but the scattering cross-sections of molecules of interest – $H_2O$, $O_2$ primarily – are high and the spatial resolution of HST is near its highest. The diffraction limit of HST at 150 nm is $\lambda/D \approx 13$ millarcseconds (mas), where $\lambda$ is the wavelength and D is the HST diameter. Although HST point spread function (PSF) at these wavelengths is significantly degraded relative to pure diffraction limited performance (see below), the images retain information at this very high spatial resolution. At a distance of 4 AU, 13 mas corresponds to 38 km, which we exploited by acquiring data in the high resolution time-tag imaging mode of the Space Telescope Imaging Spectrograph (STIS)

## 2. OBSERVATIONS

Europa was observed with the FUV multi-anode microchannel array (MAMA) on STIS using time-tag imaging mode, and the F25SRF2 filter which excludes geocoronal Lyman $\alpha$. The effective wavelength of this filter is ~150 nm, although there is a red leak discussed below. The time-tag imaging mode provides a position and time for every detected photon event, with a time resolution of 125 $\mu$s. For time-tag observations, the STIS memory is divided into two 8MB buffers, each of which can hold up to $2 \times 10^6$ events. If the cadence between scheduled buffer dumps is at least 99 seconds, then one buffer can be actively recording new events, while previously recorded events in the other buffer are being written to an HST onboard data recorder. However, in cases where Jupiter filled a large fraction of the detector, the photon event rate was such that the buffer filled in less than 99 sec, and there was a pause in the data acquisition while the buffer was read into data storage, resulting in gaps in the data acquisition sequences. We typically acquired ~40–50×$10^6$ photon events per image. Table 1 presents the exposure time for which active data acquisition occurred.

The MAMA detectors record time-tag data in "highres" mode, a 2048×2048 x-y coordinate frame with twice the spatial sampling of a standard "accum" image. The time-tag pixel size is approximately 12 mas (Biretta et al. 2016, §11.1.2).

The primary HST tracking was centered on Europa, with an additional moving target "Level 3" drift imposed to move the location of the Europa image across the detector during the course of the observation. This was to minimize artifacts arising from fine-structure irregularities in the detector sensitivity. The time-tag approach allowed us to back-out the motion of Europa to reconstruct images in the Europa (or other) coordinate frames by applying a coordinate transformation to the x-y values in the time-tag datafile, without the necessity for any spatial resampling.

We observed ten transits of Europa across the face of Jupiter spanning over a year. We also acquired seven images out of transit with the geometry illustrated in Fig. 1. The primary purpose of the out of transit images was to enable us to accurately model the appearance of Europa in the FUV, crucial in trying to model the transits themselves. The observations were designed to be taken at different telescope roll angles, but any one observation, including the acquisition procedure, is at the same roll angle and did not change throughout the exposure. Data accumulation begins once the guide star acquisition is complete. All transit observations have essentially the same viewing perspective by necessity since Europa is tidally locked in its orbit about Jupiter.

## 3. METHODS

### 3.1 ASSEMBLING THE IMAGES

Each observation resulted in a single file of photon events, which we expanded to include additional information such as new coordinate systems, tracking information, sky levels, the detector pixel flat-field value, and the location of Europa as a function of time. We refer to these expanded event files as "basetag" files. A transit image of Europa is derived by transforming the $(x,y,t)$ values for each photon event into a frame in which Europa is at rest. We chose two primary coordinate systems: one fixed in detector coordinates, with the $x,y$ values corrected for geometric distortion of the STIS FUV MAMA, and the Level 3 moving target drift removed, and the other, a frame in which Europa is oriented with the Europa North pole up and the spatial scale fixed at 35 km per pixel. These "35km" images were the principal images with which we worked. In the 35 km pixel frame, the radius of Europa is 44.6 pixels. We emphasize that because the data are time-tag, no image resampling is involved in switching between coordinate systems: the different coordinates for each photon are stored as different columns in the basetag file (there is a small contribution to the noise from digitization of the 12 mas pixel scale, negligible in our statistical analysis which uses binned data).

The task of accumulating images from the time-tag data involves first choosing a spatial coordinate system, based on a transformation of the original set of $(x,y,t)$ events, and secondly, realizing that an image is a histogram of the index value corresponding to location $(x,y)$ in the chosen coordinate frame. That is, if there are N events with unique coordinates $(x,y)$, then the image intensity value is N at location $(x,y)$. A subtlety arises when we additionally include the STIS flat-field response. Each photon event occurs at a known location on the detector, and the detector pixel flat field response ("p-flat"), or relative sensitivity, of that location is known. If the photon is detected in a region of low sensitivity, it needs to be given higher weight than a photon that arrives in a region of higher sensitivity. To do this, if the pixel p-flat value is $f$, then instead of counting "1" per event, we count "$1/f$" per event when accumulating the image from the time-tag file. This is equivalent to applying a standard flat field to the data in the case where there is no target motion or drift. To allow for various tracking solutions, we imbedded the photon events into images with pixel dimensions 4096×4096, double the size of the raw data coordinate system.

To create images in the reference frame of Europa, we compared the HST tracking solution available from the HST jitter files to the HORIZONS ephemeris. Using these two pieces of information, we predicted the $(x,y)$ position of the Europa image in the geometrically corrected detector frame. In doing so, we noticed that a small error in the HST tracking was present – the track of Europa on the detector had a curvature resulting in deviations of order one pixel for some of the data, less for the majority. Since we are working with time-tag data, this was completely removed by a coordinate transformation that followed the track of the Europa image accurately on the detector. This procedure provided a relative accuracy for the Europa location within a single observation of significantly better than a pixel. The position of the Europa image in geometrically corrected $(x,y)$ coordinates was recorded as a polynomial in the basetag header.

The absolute position of Europa on the detector is subject to uncertainties in the guide star catalog, which can cause a shift of the image away from the nominal location by a few tenths of an arcsec. Hence, having obtained an image of Europa in corrected detector coordinates, nominally located at the image center, we measured its actual position in two ways. For the transit images the most effective procedure was to cross-correlate the images with a model of the image (§3.2), and then we adjusted the position slightly by eye if necessary so that any residuals in comparison to the model appeared symmetric. By comparing the cross-correlation results to estimates done by eye, we found the image centering was good to approximately one pixel. Shifts of the model relative to the data by one pixel introduce noticeable asymmetries in their ratio. For out of transit images, we contoured the image with axis ratio and center as free parameters at a relatively low level compared to the peak count rate. This gives a highly precise measurement, but nevertheless could be slightly biased by an asymmetric light distribution. We found a consistent image elongation of ~2%, which we attributed to the convolution of the asymmetric light distribution with the complex FUV PSF. We considered and rejected centroiding, since the trailing hemisphere is approximately a factor of two darker than the leading hemisphere; this biases the centroid towards the leading edge by several pixels.

After accumulating a Europa transit image, relatively low level band structures in the background image of Jupiter remain. To estimate the optical depth of any off-limb features of Europa, we divided the transit image by an empirically constructed model of the background Jovian light. In the FUV, the face of Jupiter in the equatorial regions is relatively smooth to begin with. Perpendicular to the Jovian cloud belts, the brightness variation had an RMS brightness variation of $\approx 5\%$, while along the belts, the dispersion was consistent with Poisson noise. In the frame in which Europa is at rest, additional smoothing took place as Jupiter rotated behind Europa, smearing the Jupiter image relative to the Europa one. To model the appearance of Jupiter for the transit observation, we identified a point on the Jovian surface roughly centered on Europa during the transit, and then constructed an image using coordinate transformations in which that point on the Jovian surface is at rest. In this frame, Europa drifted across the field of view during the observation. We then built a series of images in the Jovian frame each of duration 1 sec, and masked out the data within a circular region with diameter 1.2× the diameter of Europa. We also created an exposure time image, allowing for time intervals where data acquisition was paused due to buffer dumps, and omitted the path of Europa across Jupiter. Dividing the summed Jupiter images by the exposure time image, we obtained an image of Jupiter free of the presence of Europa, which we used as the basis of an empirical model image of Jupiter.

To determine the appearance of Jupiter in each *Europa-frame* transit image, we created an image stack with 1 second slices in the Europa rest frame, using the Jupiter countrate image shifted according to the relative motion of Europa and Jupiter. We set buffer dump periods to zero, but otherwise used the entire Jupiter image. The sum of the image slices, divided by the equivalent exposure time yielded a model of the smeared Jovian background for comparison to the transit observation. After division by this empirical model, the belt structure was not present at a level above the Poisson noise. Since these Jovian models, and the actual Jovian images, were smeared in the Europa frame due to the relative motions of Europa and Jupiter, and because they were intrinsically smooth relative to the noise level of the data, we concluded that they were not introducing fine structure around the Europa limb.

This process worked well for the majority of transit observations, though there are some caveats for those observations near the Jovian limb. The relative motion of Europa and the background Jovian cloud tops is dominated by Jupiter's rotation. Near the limb, however, the apparent movement due to Jupiter's rotation decreases, and additionally the surface brightness of a location on Jupiter changes as it moves away from, or towards, the limb. The quality of the resulting models was somewhat inferior in these cases, although we were able to make the necessary models for all observations.

For all images, we measured the sky background in regions away from both Jupiter and Europa. The sky brightness was measured at 200 second intervals through the orbit, and a total background sky level was subtracted from the images prior to processing (though included appropriately in the noise models). There are strong variations in the sky background as HST approaches the Earth limb.

3.2 MODELING THE OBSERVATIONS

To assess the likelihood of image artifacts arising due to the combination of albedo fine structure on the surface of Europa, and the optics of HST, we generated a series of models of Europa and convolved them with representations of the HST PSF. The phase angle of Europa varied from ≈3° to ≈11° which had a significant influence on the illumination pattern, which we also modeled.

Qualitatively, the Europa disk appearance in the FUV images is similar to the optical, with the same large scale overall albedo variations: a dark trailing hemisphere and a bright leading one. We did not see the albedo inversion shown in Figure 4 of McGrath et al. (2009), described in more detail by Roth et al. (2014a), which is not unexpected since that appears at shorter wavelength.

The Europa model was derived from the 500 m resolution Galileo mosaic available from the US Geological Survey[1], which we mapped onto a sphere matched to the observing configuration for each visit. We neglected rotation of Europa during an observation. To allow for differences between the FUV and visible morphology, we adjusted the "contrast" of the Galileo image by a factor $C$ such that $G' = C(G-<G>)+<G>$, where $G$ is the original image and $<G>$, its mean. These contrast adjusted Galileo images were then multiplied by an illumination function, derived from the known geometry of the observation relative to the Sun, and the generalized Lambertian bidirectional reflectance distribution function (BRDF) described in Oren & Nayar (1994). This reflectance model is physically motivated, based on a macroscopic distribution of Lambertian facets (length scale >> wavelength), similar to the approach of Torrance & Sparrow (1967) for specular facets. The model is parameterized by its "roughness", which is the standard deviation of the angle distribution of facets. Rougher models have a flatter surface brightness profile, with sharper edges.

We tested a variety of PSFs: empirical and theoretically modeled using the TinyTim software package (Krist et al., 2011). The core of the PSF retains the diffraction limited width, but as aberrations increase relative to the wavelength, energy is redistributed from the core into Airy rings at greater radii. Hence, images retain the fine detail of the diffraction limit, but overlaid on that is a blurring from the broad wings of the PSF. We combined a PSF generated for the default focus position and STIS F25SRF2 filter parameters covering 15 wavelengths in the range 125 nm to 185 nm, which does not include a red leak element, with a 300 nm PSF to mimic the red leak, in the proportion 66% to 34%, which we estimated to be the proportion of counts from the Jovian reflected light spectrum bluewards and redwards of 200 nm respectively, to approximately model the red leak. *The actual value of the red leak is very uncertain*. Empirically, the smooth appearance of Jupiter, and the fact that the ice of Europa appeared darker than the clouds of Jupiter, argues that we were at least dominated by FUV photons in the images, consistent with this red leak estimate. At optical wavelengths, the clouds of Jupiter exhibit much higher contrast and the albedo of water ice (on Europa) is extremely high.

---

[1] http://astrogeology.usgs.gov/search/details/Europa/Voyager-Galileo/Europa_Voyager_GalileoSSI_global_mosaic_500m/cub

Encircled energy data for the STIS F25SRF2 PSF are available in the STIS Instrument Handbook (Biretta et al., 2016, see imaging reference material). The residual HST RMS wavefront error is in the range 25−70 nm, depending on field location, wavelength, focus, individual instrument optics, etc. (Krist & Burrows 1995; Hartig priv. comm.), which for a low 25 nm error implies a Strehl ratio (ratio of PSF peak to the theoretical diffraction limit) ≈33% at 150 nm or an encircled energy 28% of the total within the first Airy ring, radius 16 mas. The aberrations used by TinyTim yield a lower Strehl: for our composite PSF the encircled energy within a fiducial radius of 35 mas (~5 pixel diameter) is ≈30%. The encircled energy within this radius for the red leak element is 39% since the amplitude of the aberrations relative to the wavelength of observation declines to the red. We compared radial profiles of models of two of the out of transit Europa images convolved with the composite PSF and the individual red and blue components, and found that the data match more closely to the convolution with the red (sharper) PSF, and that the composite PSF is slightly too pessimistic. This could be due to either a small mismatch between the assumed aberration model of TinyTim and the actual HST optics, or a higher fraction of red leak for the icy surface of Europa than assumed. Hence, there is significant energy available at resolutions at or close to the telescope diffraction limit. Quantitatively, the amplitude of fine structure on these scales is diminished by approximately the fraction of energy in the PSF core ≈0.3−0.4, which we correct for in §4 Results, below.

Empirical PSFs were also obtained from STIS F25SRF2 observations of NGC6681 which contain numerous individual stellar images. Thirty-one separate observations used exposure times ranging from 110 to 2600 seconds. Stars were selected to be in the area of the detector where the Europa data were obtained, and were ranked according to the HST focus model available at http://www.stsci.edu/hst/observatory/focus/FocusModel. We selected 10 relatively high S/N stellar images across a wide range of focus for exploration of the data, using as empirical PSFs either the one nearest in focus to the Europa observation, or the average of the PSFs within the focus range of the Europa observation. In the end, the empirical PSF proved too diffuse in comparison to the actual data, due to resampling necessary in their preparation, coupled to inadequate S/N. Hence, although it is an idealization with inevitable small mismatches to the real PSF, we used the composite TinyTim PSF described above as our input for the data modeling, and assumed it was unchanged within a visit and from visit to visit in the detector coordinate frame.

For comparison to the data, for each visit, we developed a complete model of the scene allowing for the (slightly) different view of Europa, the differing illumination pattern due to the Europa phase angle, the Jovian background and the convolving PSF. For all visits individually, we multiplied a contrast enhanced Galileo image projected onto a sphere, by the illumination pattern of known geometry and chosen roughness. The transit models included an empirically determined image of Jupiter, smeared by the relative motions of Jupiter and Europa. The central surface brightness of the Europa model (R<0.5$R_E$), was matched to the data self-consistently by convolving the model images of Europa and Jupiter (with a gap for Europa) separately, to account for the effect of the PSF spilling light from Jupiter across the Europa image. For the out of transit images, we assumed the background sky was constant, tweaking the original measurement according to the residual sky measured in the range 1.5–3$R_E$. Fig. 3 illustrates the construction of the models.

For each observation, we generated a grid of models with a range of roughness in the range 0–1, a range of contrast enhancement from -1 to +3, and three PSFs: TinyTim, empirical nearest neighbor, and empirical range. We measured the reduced $\chi^2$ for "data minus model" to derive a fit for roughness and contrast, and spatially cross-correlated the data and model using the best fit parameters to adjust the position of the model relative to the data. Following this iterative step, we re-derived the reduced $\chi^2$ as a function of roughness, $\sigma$, and contrast, $C$, and found best-fit values of $\sigma=0.57\pm0.02$ and contrast $C=1.61\pm0.06$ using the Tiny Tim PSF. The average reduced $\chi^2 \approx 1.12$ with a range 0.98 to 1.32, omitting the leading and trailing hemisphere images which have strong "edge effects" around the Europa limb. Departures from unity were likely dominated by mismatches between the assumption that the visual albedo is directly proportional to the FUV, and that the PSF was well-represented by the Tiny Tim PSF. Nevertheless, the final models seemed acceptable, Fig. 4.

A value of $\sigma=0.57$, corresponding to a standard deviation of slopes of 32°, would be considered very rough. It is not our intent to determine the surface characteristics of Europa from this analysis, since alternative scattering models may apply (Goguen et al., 2010). They find a comparable root mean square slope distribution for the lunar highlands of ≈35° derived from optical observations, but show that an equally good fit to the photometry can be obtained with flat regolith, and multiple scattering between regolith particles on the microscopic scale. Here we were less concerned with the physical interpretation of the model, and were primarily concerned that the model empirically described the appearance of Europa to an acceptable degree of accuracy.

3.3 STATISTICAL METHODS AND RESULTS

For the statistical analysis to examine evidence for plumes, we worked with the transit observations. We analyzed the data in a variety of different ways. For the transit observations, the goal was to determine a fractional absorption against the smooth background light of Jupiter. Hence we worked with the data image divided by the convolved model, and for additional insights, we also inspected the data divided by a circularly symmetric azimuthally averaged version of the data itself (i.e. free of model assumptions). That is, we used ratio images that provide a measure of, or limit to, the optical depth of the Europa exosphere as Europa transits

Jupiter, together with the original images that gave observed counts for deriving Poisson statistics.

If the background light intensity (of light scattered by Jupiter) is $I_0(i,j)$ at each pixel in the image in the vicinity of Europa, and if the observed surface brightness is modified by a plume with optical depth $\tau(i,j)$, then the observed brightness will be $I_{obs}(i,j) = I_0(i,j)e^{-\tau(i,j)}$. For small optical depth $\tau$, and omitting the (i,j) subscript for clarity, $I_{obs} \approx I_0(1-\tau)$. If we bin the data using a box of side $n$ pixels, i.e. a total of $N=n^2$ pixels, then to obtain the average optical depth $\langle\tau\rangle$ we take one minus the average of the ratio of observed flux to background flux $\langle\tau\rangle = 1 - \langle I_{obs}/I_0\rangle$ (as opposed to $1 - <I_{obs}>/<I_0>$ which would give a luminosity weighted optical depth estimate, biasing the measurement to low values of $\tau$). An exact estimate of the optical depth is $\tau = -\ln(I_{obs}/I_0)$ however use of this quantity in the statistical comparisons would unnecessarily introduce a complex error distribution: the errors on $I_{obs}$ are straightforwardly and correctly estimated from Poisson statistics. A first order correction $\tau'$ to the optical depth may be obtained if $\langle\tau\rangle = 1 - \langle I_{obs}/I_0\rangle$ as $\tau' = \langle\tau\rangle(1 + \langle\tau\rangle/2)$.

Our initial goal was, for each image, to test whether the data are statistically consistent with the observation model described above, given the Poisson noise statistics appropriate to the image. The sky-subtracted data images in 35 km coordinate space were therefore divided by the model of the observation. The model included a reconstruction of the appearance of Jupiter, which flattened the off-limb zone, and yielded the optical depth estimate. Since we were testing for departures of the data $I_{obs}$ from the model image $I_0$, we used the noiseless model image, with the addition of the measured sky pedestal, as the basis of estimate for the Poisson count uncertainty. We tested the significance of the departure of the ratio data-to-model by calculating an estimate of the uncertainty $\sigma$ on the average $\langle I_{obs}/I_0\rangle$. The uncertainty on $\langle I_{obs}/I_0\rangle$ is given by the assumption that the uncertainty on an individual pixel is $\sigma^2(I_{obs})=(I_0+S)$ where $S$ is the sky level per pixel. Then it follows that within a box of side $n$ pixels the uncertainty on the mean within the box $\sigma(\langle I_{obs}/I_0\rangle) = \frac{1}{n}\sqrt{\langle 1/I_0\rangle + S.\langle 1/I_0^2\rangle}$. Hence we derived a $z$ statistic $z = (\langle I_{obs}/I_0\rangle - 1)/\sigma$ which we assumed was normally distributed to test for departures from a ratio value of one, i.e. optical depth zero. For the transit observations, $I_0$ was typically 15–20 counts per pixel, so the Poisson distribution was adequately approximated by the normal distribution, even for no binning. It was very well approximated by the normal distribution for modest amounts of binning (counts in the hundreds), see Table 2. We excluded pixels interior to the limb of Europa in the binning of points close to the limb.

To test the assumption of Poisson (normal) statistics, we established a sparse grid with sampling points spaced by the binning box size plus one, in the region of the image between 2.0 and 4.0 $R_E$. The means and standard deviations of the $z$ statistics for these grids are reported in Table 2.

4. RESULTS

Figures 5–14 show the results of the analysis. These images are the 512×512 pixels centered on Europa in the 35 km frame, with the Europa North pole up. On the left, we show the data image divided by the model image. A circle indicates the limb of Europa. On the right there is an image of the probability that the data are consistent with the model for the off-limb region (we blanked out the Europa disk as there were known mismatches between the model and data that were statistically significant, but which are not relevant to the off-limb transmission study). The grey scale on the left hand image ranges from 0.8 to 1.2, i.e. black would correspond to 20% decrease in $<I/I_0>$, or $\tau=0.22$. If the probability is $p$ that the data and model are consistent, for a one-sided test, then the image on the right hand side is an image of $-\log_{10}(p)$ scaled from zero to 5.5, i.e. $p$ range 1 to $10^{-5.5} \approx 3.2 \times 10^{-6}$.

For 512×512 random data points, we expect one outlier at $z \approx 4.47\,\sigma$, corresponding to a probability of $p \approx 3.8 \times 10^{-6}$, i.e. $-\log_{10} p \approx 5.4$. For a 5×5 binning factor, the number of independent data points in the image is reduced by a factor 25× and we expect to see one outlier for $z \approx 3.7\,\sigma$, corresponding to a probability of $0.95 \times 10^{-4}$, i.e. $-\log_{10} p \approx 4.0$ (as expected for approximately 100×100 independent data points). There are three images which have off-limb data points within 1.25× the radius of Europa whose $z$, or equivalently $-\log_{10} p$, values exceed this threshold. These are the transit images taken on January 26, 2014, March 17, 2014 and April 4, 2014 (see Fig. 15 and Fig. 16). The other images do not have outlier candidates by this criterion. We now look at each of these images in turn.

At face value, inspecting the January 26, 2014 image, oc7u02g2q, there are three patches of absorption in the same latitude zone that Roth et al. (2014a) found evidence of an emission line plume, though with sub-longitude $\approx 181°$ rather than the $\approx 93°$ of Roth et al. (2014a), Figs. 15 and 21 below. If we identify the plume region as -40°<latitude<-60°, radius 1.0–1.25 $R_E$, the maximum $z$-value for the ratio of the data to the model is $z \approx 3.9$ for 5×5 binning, and $z \approx 4.0$ for 7×7 binning in this region, i.e. a $\approx 4\,\sigma$ significance, and a formal probability of chance occurrence $\approx 4 \times 10^{-5}$.

Another way to look at this probability is to consider a search region close to the Europa limb. If we take an annulus of width 6 pixels (210 km), the annulus contains 1681 pixels, or 67 5×5 bins, ~34 7x7 regions. Hence the probability of obtaining a random fluctuation as large as observed with 67 or 34 trials (i.e. within the annulus close to the Europa limb) given a probability of $\approx 4 \times 10^{-5}$ per trial is only $\approx 0.0027$ (0.3%) or $\approx 0.0014$ (0.1%) for the two bin factors. These would both generally be considered statistically significant.

Independent of the models, the basic unprocessed transit image shows dark spikes starting at the limb of Europa around latitude $\approx -60°$ and extending, in the most prominent case, to a height of $\approx 6$ pixels or 220 km. Fig. 17 shows both the 35 km space (where the bands of Jupiter are horizontal) and detector space (where the bands of Jupiter are at an angle of ~60°) versions of the transit image prior to dividing by any models or images of Jupiter.

To formalize a model independent statistic, we divided the data by a circularly symmetric azimuthally averaged image, generated from the original data after division by the Jupiter model, and computed a probability following the procedures described earlier, see Fig. 18. The probability image indicates that the dark features remain statistically significant at a slightly higher significance, $z \approx 4.1$ for 5×5 binning, and $z \approx 4.8$ for 7×7 binning, in this model independent approach, and that we are not introducing artifacts during the modeling process. Further, it may be noted that the dark spikes do not appear adjacent to the darkest part of the trailing hemisphere, and they appear to have a sharp edge at the limb suggesting an origin, if real, slightly over the horizon.

As another control, the observation model with Poisson noise added was divided by the model of Jupiter to flatten out the band structure prior to azimuthal averaging. The noisy model was then divided by its azimuthal average to reveal whether spurious off-limb features appear. No off-limb features were apparent.

The second dataset we examine is ochz03dwq obtained on March 17, 2014. There is a dark spot at latitude 16.5°S, very close to the limb on the Western side but apparently protruding slightly from the limb, longitude ~275°. The protrusion is also visible in the model-independent circularly symmetric profile subtracted image, Figs. 19. Formally, the data divided by the model yields a significance level of $z \approx 4.4$ for 5×5 binning, a probability of chance occurrence $p \approx 4.8 \times 10^{-6}$.

To test for a problem with the model, we divided this image (after division by the Jupiter model) by the median of the other nine transit images, treated identically, since the view of Europa in transit is always very similar. The dark spot remained in this ratio. To estimate its significance, we derived an uncertainty estimate on the ratio using standard propagation of errors, assuming that the error on the median was $\sqrt{\pi/2}$ times the error on the mean, as it is for a Gaussian distribution. Fig. 20 shows the data points in the image beyond the limb of Europa normalized by the error estimate, with the position of the dark spot indicated. Again, the protrusion appears to be statistically significant, with $z \approx 6.6$, though since this quantity is the ratio of two random variables, the distribution is not formally Gaussian, although the zero mean unit dispersion Gaussian included for comparison does appears to be quite a good description of the distribution. Empirically, the patch under discussion is the darkest spot on the image.

In addition, these data show similar darkenings off-limb in the south polar zone to the January 26, 2014 image: as for that image, there are two darker patches in the same latitude range at polar angles 236 and 260°, with $z \approx 4.0$.

Image ochz05ftq was obtained on April 4, 2014. It shows a single dark spot at latitude 40°S on the trailing hemisphere, polar angle 220°, see Figs. 9 & 15. With 5×5 binning, it is at about 4.5 σ, comparable to, or even slightly more significant than, the other candidate detections. We are concerned that the rim of apparently darker material in this image is mirrored in part by a rim of brighter material on the opposite side of Europa, seen as a dark area in the probability maps. Though not at such a strong significance level, this bright rim could indicate a problem in the data processing. This image also has a (small) number of outliers elsewhere in the field of view, Table 2.

## 5. SYSTEMATIC ERRORS AND CONCERNS

Even if the data show statistically significant absorption features according to the photon statistical analysis above, which assumes independent randomly distributed data points, it is conceivable that systematic effects may come into play to introduce artifacts and cast doubt upon the reality of the dark patches. We step through a range of potential sources of systematic error. None of these appear to be responsible for the features.

*Detector non-Poisson behavior:* If the counting statistics were not Poisson, but influenced in some fashion by the physics of the detection process or reliability of the MAMA counting methods, then there may be random fluctuations larger than the Poisson distribution would predict. Negative fluctuations of this sort could be mistaken for plumes. Table 2 shows, however, that the number of 4.5 σ outliers in the control region beyond 1.5 $R_E$ are essentially as expected from Poisson statistics. We expect one event by chance at 4.47 σ for 512x512 trials. We see zero for six of the images, 1 for two of them and 3 for two images. We conclude that there is not a tendency for the detector to exhibit anomalous dark spots in general, and for such dark spots to congregate around the Europa limb demands a different explanation. Overall, the dispersion of the normalized *z* statistics is within 2–3% of the Poisson estimate (1.8% for 5×5 binning, and 2.3% for 7×7), and the mean is close to zero 0.05–0.06, Table 2. The remaining residuals are likely due to imperfectly modeled Jupiter band structure, from inspection of Figs. 5–14. Thus from the statistics of the control region, there is no compelling evidence that the detector has non-Poisson counting behavior.

*A complex and variable point spread function:* Fig. 2 illustrated the complexity of the FUV PSF. For a root mean square wavefront error of ≈25 nm for the HST corrected optics, the estimated Strehl ratio is ≈0.33, significantly below diffraction limited performance. This manifests itself as a tight core, retaining diffraction limited information, but with substantial "power" in the wings of the PSF. To make matters worse, there is a contribution to the PSF from the only approximately known red leak, which also introduces a scene-dependence to the PSF. That is, the regions of high albedo exposed ice, the darker trailing hemisphere material and the reflected light from Jupiter each must have a slightly different PSF. Further, the HST focus changes through an observation, introducing a time dependence to the PSF. Hence it is not possible to know accurately the exact PSF for each part of the image at all times.

However, when convolving the complex PSF with the extended halo of Jovian light in the transit images, the effect of the substantial PSF wings is to smooth the resulting image, rather than to introduce sharp features such as the candidate plume observations, as discussed in §3.2.

*Detector defects:* There are a number of places on the STIS MAMA detector where defects are evident, and at high resolution, the appearance of the detector p-flat is of a honeycomb grid with significant amplitude variations through the grid lines. We addressed this in two ways. Firstly, we used our own observations of Jupiter, to self-consistently derive an improved p-flat calibration, which by definition is appropriate to the time frame of our observations. Secondly, we used the Level 3 moving target specification for HST tracking to drift the image of Europa across the detector during the observation. Hence any localized detector defects and the honeycomb grid, were smeared out over the length of the drift. After gaining experience of these issues, we adopted a primary pointing of Europa to be well away from detector defects. The dark features in question (putative plumes) are significantly more localized than Level 3 smeared detector defects, and they are in a different position angle to the imposed tracking drift. We inspected the effective p-flat for each observation and did not see any defects coincident with candidate absorption features.

*Model mismatches to the data:* The comparison model is not perfect – we fitted for contrast and illumination function, and for the location of Europa within the image as described, and derived the underlying image of Jupiter empirically and self-consistently from the data. We checked for the effect of centering errors by eye on two of the candidate plume images, and found that the appearance of the image is robust against shifts of order one to two pixels, by which time the asymmetries in the ratio of data to model are noticeable. The contrast value and illumination functions were fitted by least squares and have relatively small errors, quoted above. Small departures from these fitted parameters do not affect the appearance of off-limb structure in comparison with the data: we tried several version of the fits depending on which PSF was used and found no substantial difference in the appearance of the ratio images.

*There are additional reasons to be cautious:*

The first image that shows apparent off-limb structure, oc7u02g2q, used the smallest Level 3 drift which means detector features are more pronounced in the flat field.

The darkest spot in ochz03dwq is close to, though not exactly coincident with the darkest part of the Europa disk. It is also apparent on the image with the largest sublongitude, indicating we are seeing further "around" the dark trailing hemisphere by ~1.5 degrees. Equivalently, this image is the one with Europa closest to the Jovian limb, making correction for the Jovian background harder.

The individual best optical depth (1-$\sigma$) limit is 0.033 for 7×7 binning in oc7u02g2q. While this may seem to be good, if there are weak systematics, the higher sensitivity could render them more statistically significant.

All features are in roughly, but not exactly, the same area of Europa – a strength yielding insight into the physics of Europa, or an indication of an unsolved imaging problem?

Despite almost contemporaneous observations with oc7u02g2q (January 26, 2014) during early 2014, Roth et al. (2014b) failed to detect any plume activity. They observed Europa twice, January 22 and February 2, 2014, indicated on Fig. 16, and derived only upper limits. The difference does not appear to be a geometric viewing perspective difference. All transit observations have subobserver longitude ≈180°, and the 2014 Roth data were acquired with subobserver longitude in the range 117–157°, which, coupled to the extreme polar latitudes involved, means the features ought to have been visible to Roth et al. (2014b). There are possible physical explanations, which we return to in the discussion, however the apparent inconsistency gives us pause, and represents a cause for concern.

6. DISCUSSION

6.1 QUANTITATIVE IMPLICATIONS

If the statistically significant regions are due to plumes, then we can quantify their properties, with the caveat that these numbers are to be treated with caution, and *apply only if the features are actually plumes and not artifacts*. Table 2 gives the mean dispersion in the bright region surrounding Europa (i.e. the transit image divided by the model of Jupiter). This may be interpreted as the "1-σ" optical depth limit, for which the average value is 0.056 and 0.04 respectively for 5×5 and 7×7 binning. The individual best limit (1-σ) is 0.033 for 7×7 binning in oc7u02g2q, which also has the second best spatial resolution and is one of the images which appears to show statistically significant off-limb features. The actual candidate features were seen at ≈4.5 σ, corresponding to a lower detection limit of τ≈0.15–0.25. There is an additional correction factor due to the broad wings of the PSF, as discussed in §3.2. This correction factor is size dependent, larger features requiring smaller correction. By simulating a grid of Gaussian patches of absorption with a range of optical depth and length scale, we found that for the length scales of interest, a factor of ≈2.3–3.3 to scale the as-seen integrated extinction over the patch, to the underlying total, was appropriate, corresponding to a PSF with ≈37% of its energy in the compact core (see §3.2). Hence, taking the mean factor 2.8× and correcting the optical depth, we derived a lowest detectable limit of intrinsic optical depth of τ≈0.4. We illustrate one of these models below, applied to the data of March 17, 2014.

To quantify the amount of water required to produce a patch of absorption, we looked at both molecular and ice particle cross-sections. The STIS FUV MAMA detector has a significant red leak. Using the component level software package *pysynphot*, with a Solar spectrum, we calculated that the fraction of detected photons with wavelength longer than 200 nm is approximately 34%. This acts to dilute the effective cross-section of absorbing molecules, typically highest in the FUV region, which represents ≈66% of detected photons. The actual value of the red leak is very uncertain – this value was obtained using pre-launch component throughput curves; it has not been calibrated on orbit or as a function of time. Qualitatively, however, the smooth appearance of Jupiter, and dark appearance of Europa indicate that the image is dominated by the FUV spectrum and not the red leak.

To obtain an effective cross section for $H_2O$, in light of these considerations, we derived a countrate-weighted mean cross section using the *pysynphot* spectrum together with the cross-section as a function of wavelength. The resulting wavelength averaged weighted cross-section was $\sigma(H_2O) \approx 1.8 \times 10^{-22} m^2$. For a patch of area $A_{35}$ pixels of 35 km size, and optical depth $\tau_v$ the corresponding implied mass in water vapor due to molecular absorption is $M_v = 0.2 \times 10^6 \tau_v A_{35}$ kg.

The plumes of Enceladus comprise a mixture of water vapor and ~micron sized ice grains (Dong et al., 2015). The ratio by mass of grains to vapor was variable, but in the range 15–25% typically. The ultraviolet scattering cross-sections for solid column (and other) ice particle habits are given by Key et al. (2002). Converting their parameters into mass and optical depth, we derive a mass in grains of $M_g(H_2O) = 0.64 \times 10^6 \tau_g A_{35} a_{\mu m}$ kg, where $\tau_g$ is the optical depth due to grains, $A_{35}$ is the area of the patch in 35 km pixels, and $a_{\mu m}$ is the grain size in microns. The total optical depth is the optical depth due to molecular absorption (vapor) and grains. If the mass ratio of grains to vapor is $\alpha \approx 0.2$, the total inferred mass in water from a patch of intrinsic optical depth $\tau$ and area $A_{35}$ pixels is $M_{tot}(H_2O) = 0.2 \times 10^6 \tau A_{35} (1+\alpha)/(1+\alpha/3.2a_{\mu m})$ kg, from which it is evident that the presence of grains in any scattering ice plumes makes a difference only of order 10% (12.9% specifically for $\alpha=0.2$). More subtly, however, is that grains scatter across the entire spectrum, including the redder regions for which the PSF is sharper, raising the possibility that the correction factor due to PSF blurring may be (slightly) reduced. In the discussion that follows, given these offsetting small corrections, we set $\alpha=0$.

At face value, inspecting the January 26, 2014 image, there are three patches of absorption in the latitude zone where Roth et al. (2014a) found evidence of an emission line plume, though with sub-longitude $\approx 181°$ rather than $\approx 93°$, Fig. 21. We defined the potential plume region as those points within the Roth latitude range (polar angle 220–260°), and probability of chance occurrence $p<0.01$ based on the probability image for 5×5 binning, for the quantitative analysis that follows.

Converting the statistical data to physical quantities, the average optical depth, defined as $I=I_0 e^{-\tau}$, in the plume region as described in the previous paragraph, is $<\tau>\approx 0.15$ (15% absorption) averaged over 45 pixels (55125 km$^2$) with a peak $\tau_{max} \approx 0.25$. For an average measured optical depth $<\tau>\approx 0.15$ the implied intrinsic optical depth is $<\tau>\approx 0.42$ for a blurring correction factor of 2.8, and the average column density is $<N>\approx 2.3 \times 10^{21}$m$^{-2}$ which is higher than the H$_2$O column density inferred by Roth et al. (2014a) of $\approx 1.5 \times 10^{20}$m$^{-2}$. Nevertheless, the total number of H$_2$O molecules in the two-plume model of Roth et al. (2014a) is $1.3 \times 10^{32}$, while the implied total number of H$_2$O molecules (column density times projected putative plume area) of our new observation is also $1.3 \times 10^{32}$, or $\approx 3.9 \times 10^6$kg, in good agreement, though the uncertainty is significant.

Hence, at face value, the absorbing features we see imply a similar order of magnitude of material to the emission features of Roth et al. (2014a), and arise in the same latitudes, but with physical characteristics derived in a completely independent fashion using a very different observing strategy, albeit with the same detector.

We turn now to the compact nub of March 17, 2014, on the trailing limb. Quantitatively, the average optical depth over the central 22 pixels is $<\tau>\approx 0.21$, reaching a peak $\tau \approx 0.34$ at a significance level of 4.4 $\sigma$, with formal probability of being a random event $\approx 4.8 \times 10^{-6}$. The corresponding column density for a pure water event would be $3.3 \times 10^{21}$m$^{-2}$ corresponding to a total of $1.8 \times 10^{32}$ H$_2$O molecules, or $\approx 5.4 \times 10^6$kg. The two polar region patches of absorption in this image have an implied corresponding column density $2.7 \times 10^{21}$m$^{-2}$ corresponding to a total of $2.1 \times 10^{32}$ H$_2$O molecules, or $\approx 6.6 \times 10^6$kg.

Fig. 22 shows an example of an idealized model, with a Gaussian $\sigma=2.5$ pixels ($\approx 90$ km) absorption patch, peak optical depth 0.75, with Poisson noise added, not intended to be a rigorous fit to the data, but in quite good qualitative and quantitative agreement with the data, also shown. In this example, the peak implied intrinsic optical depth is $\approx 2\times$ the peak observed optical depth and the total intrinsic absorption of the model is 2.3× the face-value extinction in the data.

Examining the image of April 4, 2014, the average optical depth is 0.04 over 42 pixels, reaching a peak τ≈0.3 at a significance level of 4.5 σ, with formal probability of being a random event ≈2.9×10$^{-6}$. The corresponding average column density for a pure water event would be 0.7×10$^{21}$m$^{-2}$ corresponding to a total of 0.5×10$^{32}$ H$_2$O molecules, or ≈1.4×10$^6$kg.

Given large correction factors due to pixel blurring, red leak and the balance of ice particle to molecular scattering, the numbers above are intended to provide only an approximate order of magnitude to assess plausibility. Since the total amounts of implied material are of the same order as those of Roth et al. (2014a), we conclude that it is possible water plumes could be producing the apparent absorption features.

6.2 A PHYSICAL MODEL?

The original hypothesis being tested by Roth et al. (2014a) was that, following the example of Enceladus, the plumes were modulated on the orbital period of Europa (3.55 days) and that maximum activity would occur at apoapsis. The true anomaly values for the three transits potentially indicative of activity, are intermediate between apoapsis and periapsis (see Table 1 and Fig.1). These three candidate images are grouped together covering a time period of ~2.5 months, though with a gap on March 24, 2014 (Fig. 16), hinting at a protracted period of activity. If this were to represent a timescale of activity, the two month period is ~19 orbits of Europa about Jupiter, many times the Europa orbital period (3.55 days); hence the initial expectation of an activity cycle that correlates with Europa's location in its orbit, similar to Enceladus and as proposed by Roth et al. (2014a), is not valid.

Also, despite almost contemporaneous observations during early 2014, Roth et al. (2014b) failed to detect any plume activity. There are possible physical explanations: the plumes may be genuinely transient if their activity is governed by tidal stresses as Europa orbits Jupiter. The observing programs of Roth et al. (2014a; b) were designed to test this hypothesis, and they concluded that other factors must be involved. Variations in the plasma environment cause large variations in auroral activity, and even if plume activity is fairly constant, or slowly varying, rapid fluctuations in the ambient plasma can result in substantially different detectability of the plume emission sought by Roth et al. (2014a, b). The upper limits of Roth et al. (2014a, b) are typically a factor of a few less than their detection, which in principle could be due to decreased excitation rather than diminished plume activity. Our technique is insensitive to the plasma environment.

The feature of March 17, 2014 is close to the crater Pwyll, but is significantly displaced to the North, by approximately 6.5° of latitude, or 177 km. Fig. 23 plots an ellipse of dimensions 8.3°×34° centered at longitude 275.68°, latitude 16.4°S, corresponding to the approximate uncertainty in the location of the feature from the HST image. The apparent extent of the feature along the limb is ≈6.6 pixels and we assume an uncertainty of 2 pixels orthogonal to the limb. Fagents (2003) summarizes a wide variety of models that could explain cryogenic volcanism on the surface of Europa, though acknowledging that despite the geological youth of the surface, there may be other explanations for the features considered. There is a great deal of complex terrain within the ellipse shown in Fig. 23, and several features reminiscent of those discussed in Fagents (2003) in the context of cryovolcanism. However, we consider discussion of detailed physical processes premature until the plume activity is confirmed (or shown not to exist).

7. CONCLUSIONS

We have used HST STIS FUV imaging observations of Europa as it transits the smooth face of Jupiter to reveal statistically significant evidence of off-limb absorption features in the vicinity of the plumes discovered by Roth et al. (2014a). All transit observations have essentially the same viewing perspective since the orbit of Europa is tidally locked. The features are mostly at similar latitude to the event of Roth et al. (2014a), but approximately 90° further West in longitude. There are hints that the features appear three times between January 2014 and April 2014. The original hypothesis tested by Roth et al (2014a) was that plume activity would correlate with true anomaly, as is the case for Enceladus. These times, however, do not correspond to special values of the true anomaly, in particular orbital apocenter, and Roth et al (2014b) also came to the conclusion, namely that being at orbital apocenter is not a sufficient condition for plume activity. This period also, however, encompasses the null results of Roth et al. (2014b). There are potential physical explanations, including a variable plasma environment or intermittent activity, or this may be indicative of the presence of unknown artifacts. If the apparent absorption is due to plume activity, quantitatively, the implied amounts of water are similar to the plumes described by Roth et al. (2014a).

We find only upper limits for all other observations between December 2013 and March 2015.

The observation of March 17, 2014, provides an indication of a compact off-limb patch of absorption at a more equatorial latitude $\approx 16.5°$S, longitude $\approx 275°$W, north of the crater Pwyll. The peak optical depth of the patch is $\tau \approx 0.34$, corresponding to a column density of $\approx 1.8 \times 10^{21}$ m$^{-2}$ and a mass of $\approx 5.4 \times 10^6$ kg if the feature is pure water. Since this feature is at a more equatorial latitude, and is compact, it is possible to locate the event more precisely on the surface of Europa. Even so, the error ellipse encompasses a wide variety of Europa terrain.

In conclusion, we find tentative evidence in three of our ten observations that could be indicative of plume activity on Europa. The features we find are statistically significant from the perspective of photon statistics, and we are unable to identify any definitive systematic problem that could be responsible. If the features we see are due to plumes, then the activity must be more common and more extensive than previously thought.

ACKNOWLEDGMENTS

These data were obtained using the Hubble Space Telescope which is operated by STScI/AURA under grant NAS5-26555. We acknowledge support from grants associated with observing programs HST GO-13438, HST GO/DD-13620 and HST GO-13829.

TABLES:

TABLE 1: Summary of STIS observations

| Dataset | Visit Start Time | Exposure time (seconds) | Observation sub longitude | Europa Diameter (arcsec) | km per pixel (detector) | True Anomaly |
|---|---|---|---|---|---|---|
| OC7U03SCQ* | 12/22/13 6:58 | 1473.88 | 182.12 | 1.018 | 37.7 | 238.91 |
| OC7U02G2Q* | 1/26/14 18:05 | 2023.24 | 181.75 | 1.006 | 38.1 | 260.08 |
| OCHZ09QRQ | 2/27/14 19:57 | 2507.19 | 195.65 | 0.9317 | 41.1 | 296.56 |
| OCHZ03DWQ* | 3/17/14 11:47 | 2389.45 | 185.68 | 0.8816 | 43.5 | 296.83 |
| OCHZ04EAQ* | 3/24/14 12:45 | 2232.8 | 178.98 | 0.8619 | 44.5 | 295.03 |
| OCHZ05FTQ* | 4/4/14 5:20 | 2200.41 | 181.9 | 0.8332 | 46.0 | 310.31 |
| OCHZ06EVQ* | 4/14/14 21:52 | 2119.05 | 184.3 | 0.8065 | 47.5 | 316.94 |
| OCHZ08FYQ* | 4/22/14 0:19 | 1802.89 | 183.33 | 0.7902 | 48.5 | 323.85 |
| OCHZ07P2Q* | 5/2/14 15:15 | 2507.2 | 178.55 | 0.7679 | 49.9 | 328.72 |
| OCP206JFQ | 12/28/14 14:26 | 2508.19 | 348.86 | 0.9398 | 40.8 | 354.47 |
| OCP207JWQ* | 1/6/15 14:54 | 1741.85 | 184.01 | 0.9607 | 39.9 | 194.46 |
| OCP202X3Q | 1/9/15 4:58 | 2508.19 | 86.37 | 0.9645 | 39.7 | 100.43 |
| OCP204D7Q | 1/10/15 6:26 | 2508.2 | 194.12 | 0.9674 | 39.6 | 208.49 |
| OCP203M8Q | 1/17/15 2:23 | 2508.2 | 167.36 | 0.9781 | 39.2 | 185.64 |
| OCP205N1Q | 2/15/15 11:28 | 2508.2 | 269.45 | 0.9884 | 38.8 | 306.55 |
| OCP208PBQ* | 3/4/15 7:45 | 1489.27 | 178.99 | 0.9691 | 39.5 | 227.08 |
| OCP251YTQ | 3/6/15 5:54 | 2508.17 | 14.09 | 0.9638 | 39.8 | 66.44 |

TABLE 2: Statistical summary of image quality

| Dataset | Counts | Sky (counts) | $\sigma(d/m_5)$ | $\sigma(d/m_7)$ | $<z5>$ | $\sigma(z5)$ | $<z7>$ | $\sigma(z7)$ | n5 | n7 |
|---|---|---|---|---|---|---|---|---|---|---|
| oc7u03scq | 13.68 | 0.13 | 0.054 | 0.039 | 0.009 | 1.001 | 0.048 | 0.995 | 0 | 0 |
| oc7u02g2q | 19.11 | 0.21 | 0.046 | 0.033 | -0.008 | 1.015 | -0.005 | 1.025 | 0 | 0 |
| ochz03dwq | 16.23 | 0.6 | 0.05 | 0.035 | 0.063 | 0.983 | 0.084 | 0.988 | 0 | 0 |
| ochz04eaq | 13.12 | 0.81 | 0.057 | 0.042 | 0.072 | 1.011 | 0.074 | 1.030 | 1 | 0 |
| ochz05ftq | 12.43 | 0.52 | 0.059 | 0.042 | 0.052 | 1.026 | 0.046 | 1.029 | 3 | 2 |
| ochz06evq | 13.54 | 0.48 | 0.057 | 0.04 | -0.004 | 1.027 | 0.070 | 1.013 | 0 | 2 |
| ochz08fyq | 12.47 | 0.59 | 0.059 | 0.042 | 0.030 | 1.015 | 0.038 | 1.016 | 0 | 0 |
| ochz07p2q | 9.8 | 0.52 | 0.068 | 0.049 | 0.055 | 1.040 | 0.083 | 1.051 | 1 | 1 |
| ocp207jwq | 13.85 | 0.23 | 0.054 | 0.039 | 0.071 | 1.017 | 0.130 | 1.038 | 0 | 0 |
| ocp208pbq | 15.04 | 0.08 | 0.053 | 0.038 | 0.052 | 1.047 | 0.059 | 1.040 | 3 | 5 |
| Mean | | | 0.056 | 0.040 | 0.039 | 1.018 | 0.063 | 1.023 | 0.8 | 1 |

Columns:
1. Dataset
2. Mean counts outside 1.5 $R_E$ in 35 km frame
3. Sky level in 35 km image
4. Standard deviation of data/model in sampling grid, yielding 1-$\sigma$ limit for optical depth, bin 5×5
5. Standard deviation of data/model in sampling grid, yielding 1-$\sigma$ limit for optical depth, bin 7×7
6. Mean normalized data/model (i.e. "*z*") in sampling grid, bin 5×5
7. Standard deviation of normalized data/model in sampling grid, bin 5×5
8. Mean normalized data/model (i.e. "*z*") in sampling grid, bin 7×7
9. Standard deviation of normalized data/model in sampling grid, bin 7×7
10. Number of negative outliers, 5×5 binning
11. Number of negative outliers, 7×7 binning

FIGURE CAPTIONS

Figure 1. Left, true anomaly distribution for HST STIS observations. Transits are in blue and out of transit, red. The transits with evidence of plume activity are circled. Right, the configuration of Europa as seen by the observer.

Fig. 2. Left, central 2.5×2.5 arcsec of the composite TinyTim model PSF used, and for comparison, empirical PSFs (images of stars) at different focus values illustrating the difficulty of exactly matching the model PSF to the data.

Fig. 3. Procedure used to construct model observations. An optical map of Europa was combined with a generalized Lambertian illumination function and background reconstruction of Jupiter, convolved with the PSF, central 2.5 arcsec shown, and Poisson noise added.

Fig. 4. Lower array shows all images of Europa in geometrically corrected detector coordinate system, ranked in time starting lower left, moving left to right. Upper panel shows derived models for each observation, without Poisson noise.

Figures 5-14: The following 10 figures show data/model on the left hand side, scaled 0.8-1.2 corresponding to 20% absorption as black, and on the right, the probability of chance occurrence, given by ($-\log_{10}p$) scaled 0–5.5 with high numbers representing low probability of chance occurrence.

Fig. 5. oc7u03scq  
Fig. 6. oc7u02g2q  
Fig. 7. ochz03dwq  
Fig. 8. ochz04eaq  
Fig. 9. ochz05ftq  
Fig. 10. ochz06evq  
Fig. 11. ochz08fyq  
Fig. 12. ochz07p2q  
Fig. 13. ocp207jwq  
Fig. 14. ocp208pbq  

Fig. 15. Probability images for Europa transits on (a) January 26, (b) March 17, (c) April 4, 2014, showing statistically significant off-limb features, based purely on photon statistics without regard for possible systematic effects. Beneath, (d), (e), (f), the corresponding "optical depth" images scaled 0.0–0.15 black to white. Inset to lower right shows the PSF to the same scale with the same bin factor as the images. Images: oc7u02g2q, ochz03dwq, ochz05ftq.

Fig. 16. Timeline for Europa transit observations, showing significant residuals between January and April 2014. Images are (-log) probability images, as for Figs. 5 to 14, scaled 0–5.5.

Fig. 17. Image oc7u02g2q obtained January 26, 2014 shown in two coordinate frames: 35 km/pixel (left) and detector frame (right). The features of interest are indicated with the ellipse.

Fig. 18. Image of oc7u02g2q after division by a circularly symmetric profile, left, with associated probability map on the right, showing the same features as seen in the data/model image.

Fig. 19, showing dark patch close to limb in image ochz03dwq. Top left, the data image after dividing by a model of Jupiter; lower left data after division by circularly symmetric profile. Top right, data divided by model, and bottom right, data divided by median of the other 9 transits. Dashed green circle shows Europa limb, and blue circle centered on dark spot.

Fig. 20. Histogram of normalized deviations "$z$ values" for ochz03dwq, divided by the median of the other 9 transit images processed as described in the text. The location of the minimum z of the compact protrusion discussed in the text is indicated with the blue arrow. A Gaussian with zero mean and unit standard deviation is overplotted.

Fig. 21. Showing the location of features in the January 26, 2014 image relative to the location of the Roth et al. (2014) plume detection. The cyan circle shows the new location and the yellow, the Roth one. On the right, the Roth et al. (2014) annulus and sector containing the putative plumes, which also contains the current features.

Fig. 22. Illustration of a model patch of absorption, convolved with a TinyTim PSF and Possion noise added, configured to approximately match the data of March 17, 2014. The model is to the left and the data, to the right.

Fig. 23. High resolution image from Galileo showing approximate location uncertainty of the equatorial off-limb feature discussed in the text. The viewing perspective of the STIS data is from the right.

Figures

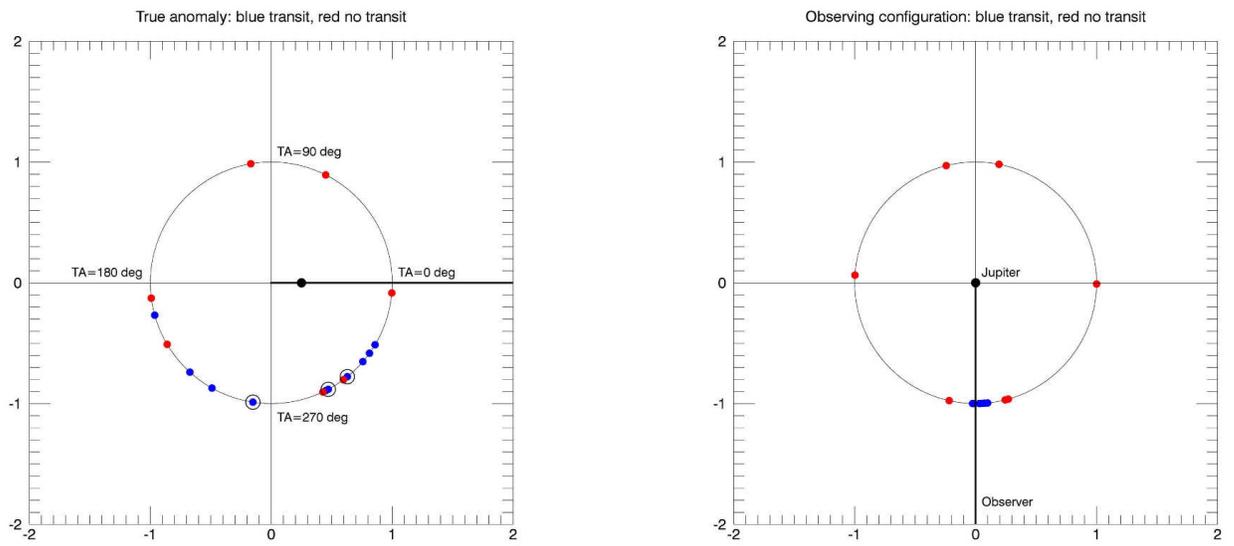

Figure 1

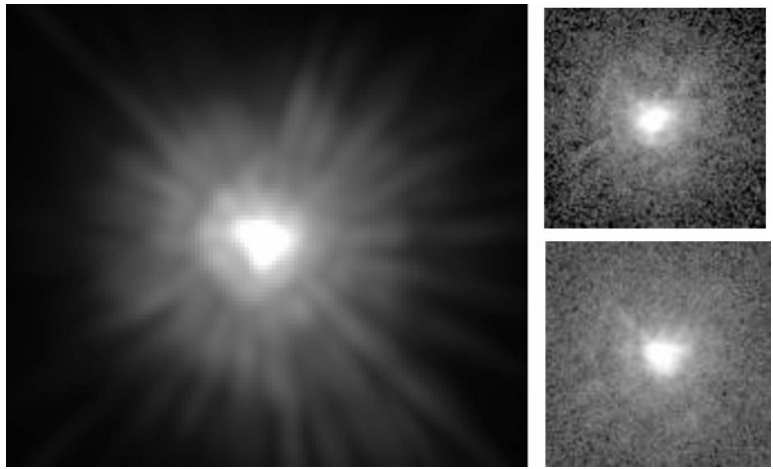

Figure 2

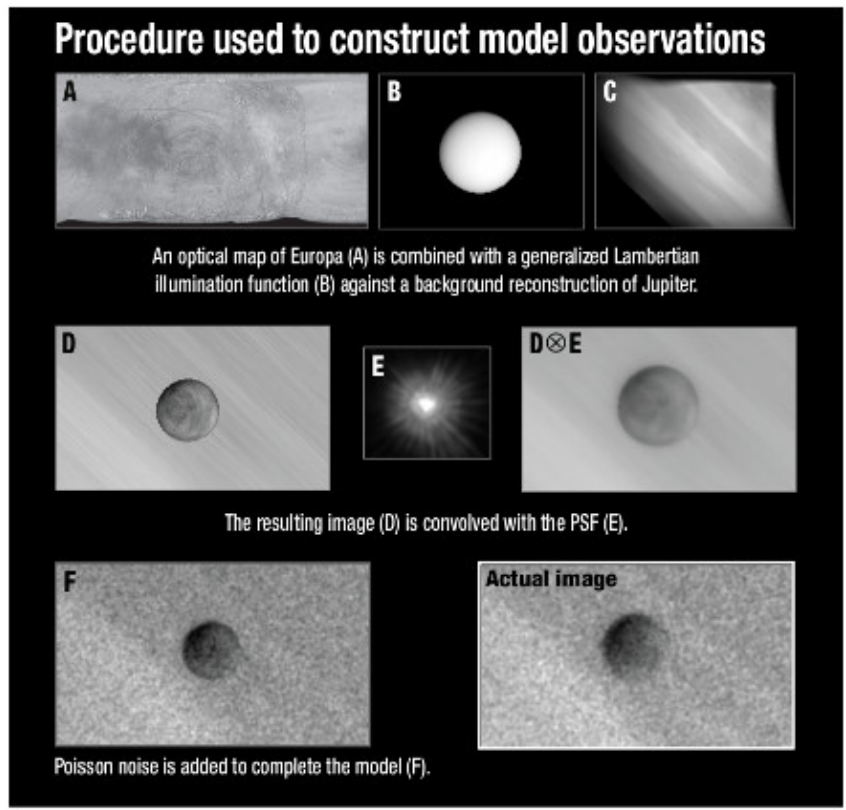

Figure 3

**Modeled data**

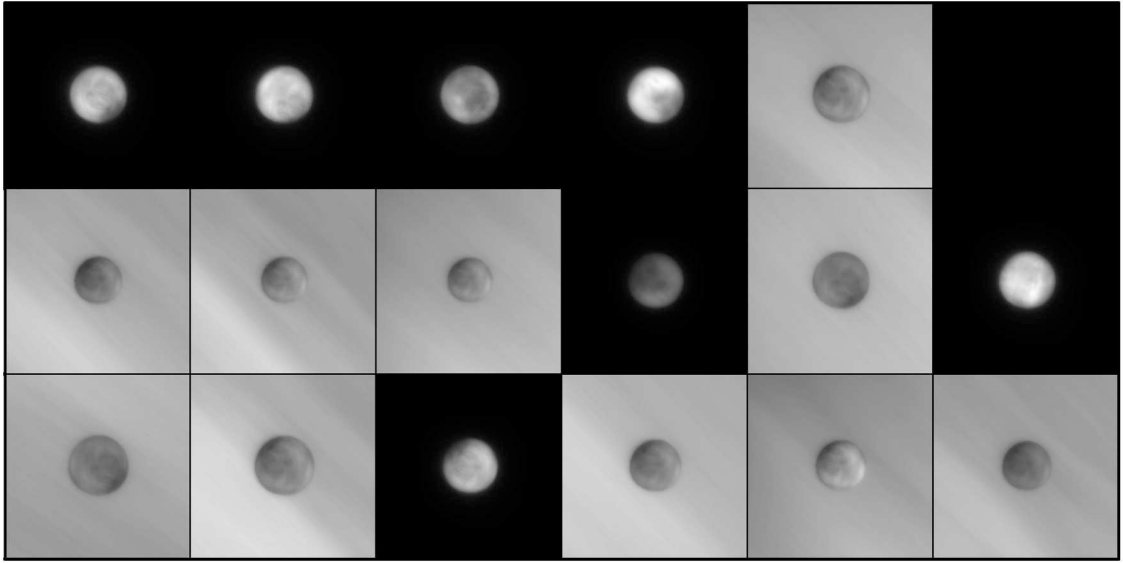

**Actual data**

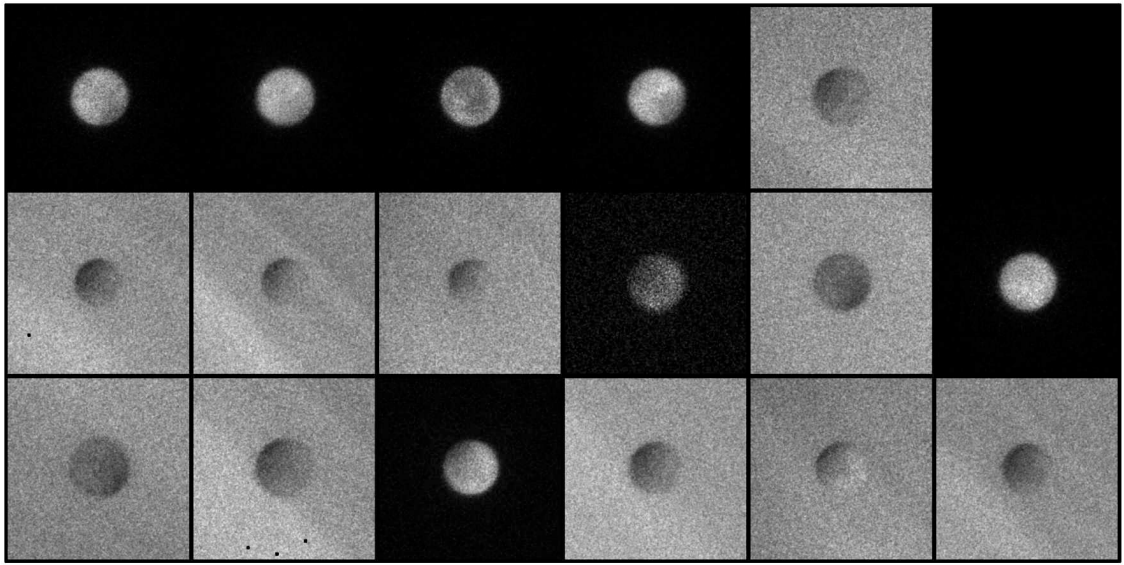

Figure 4

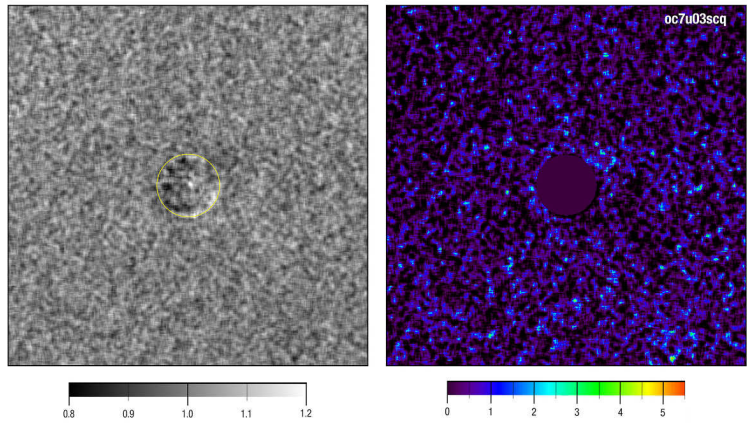
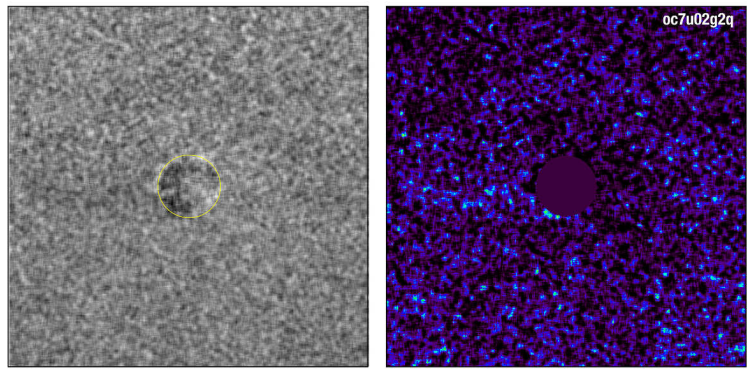
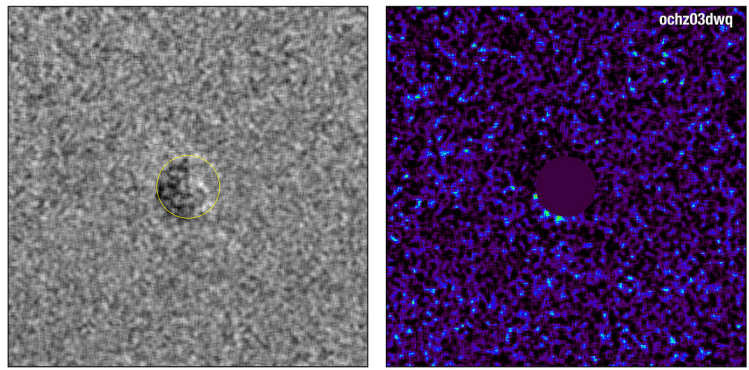

Figures 5, 6 and 7

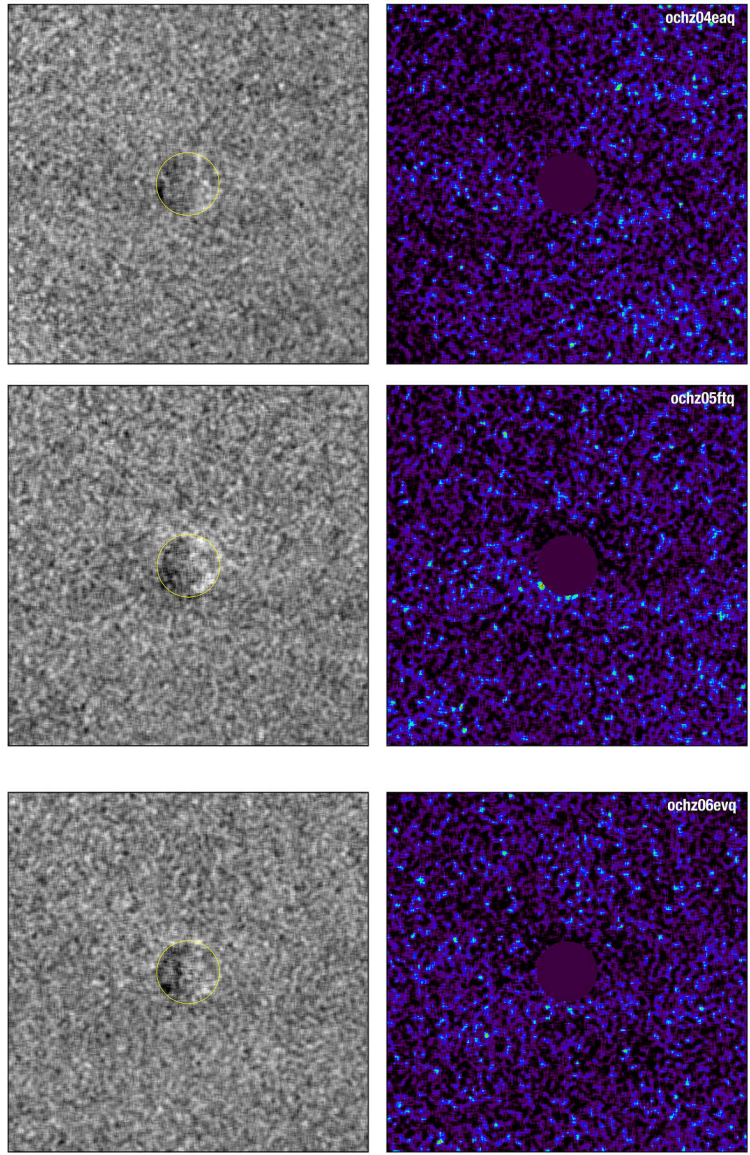

Figures 8, 9 and 10

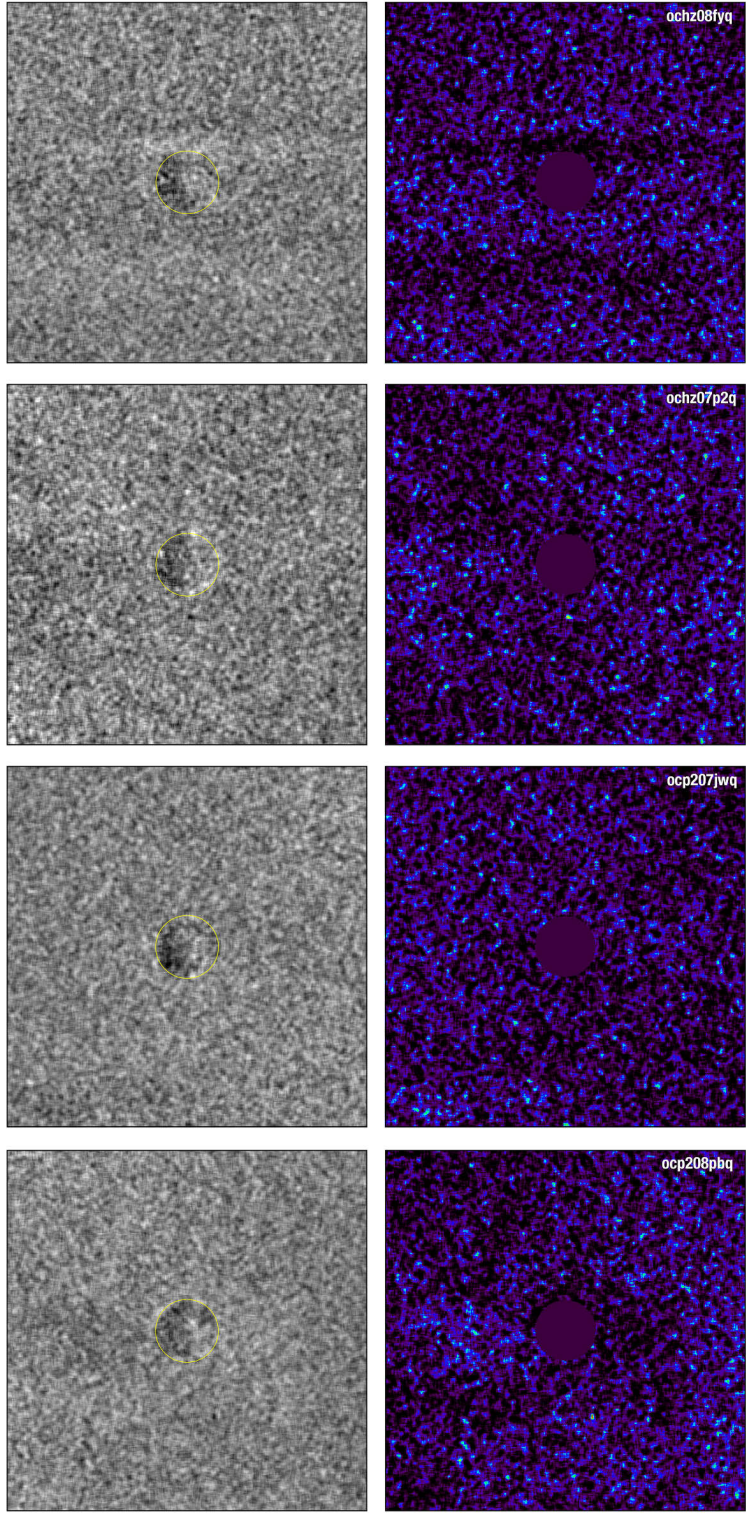

Figures 11, 12 13 and 14

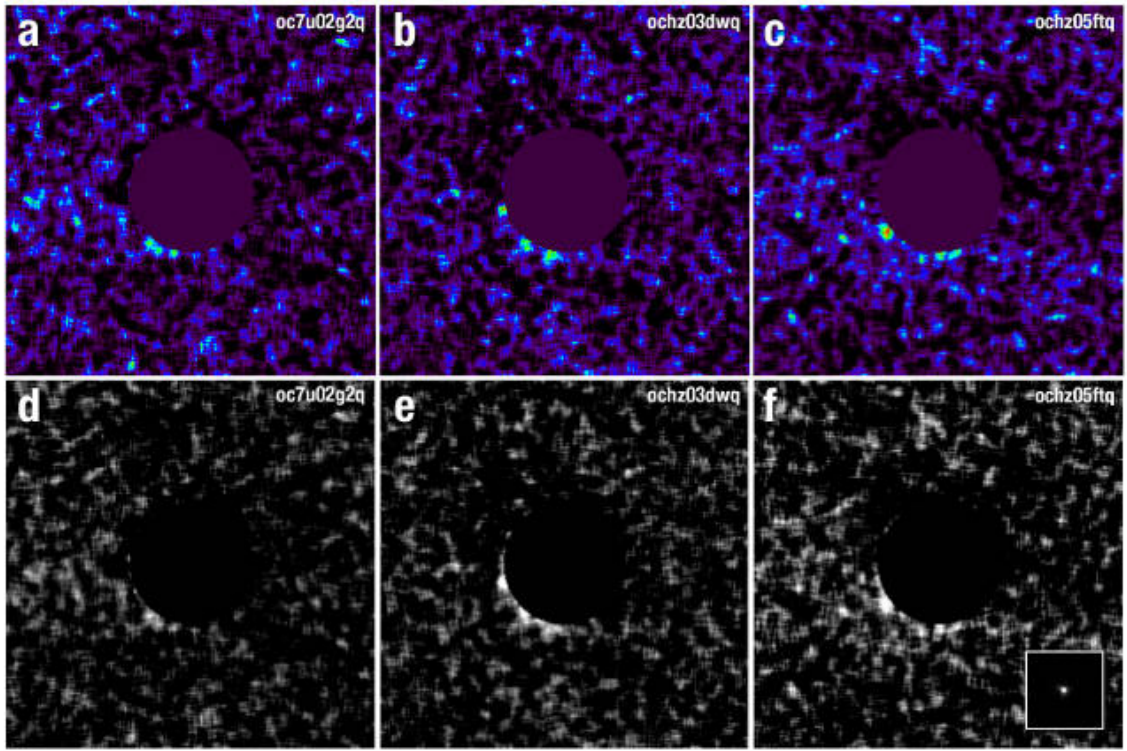

Figure 15

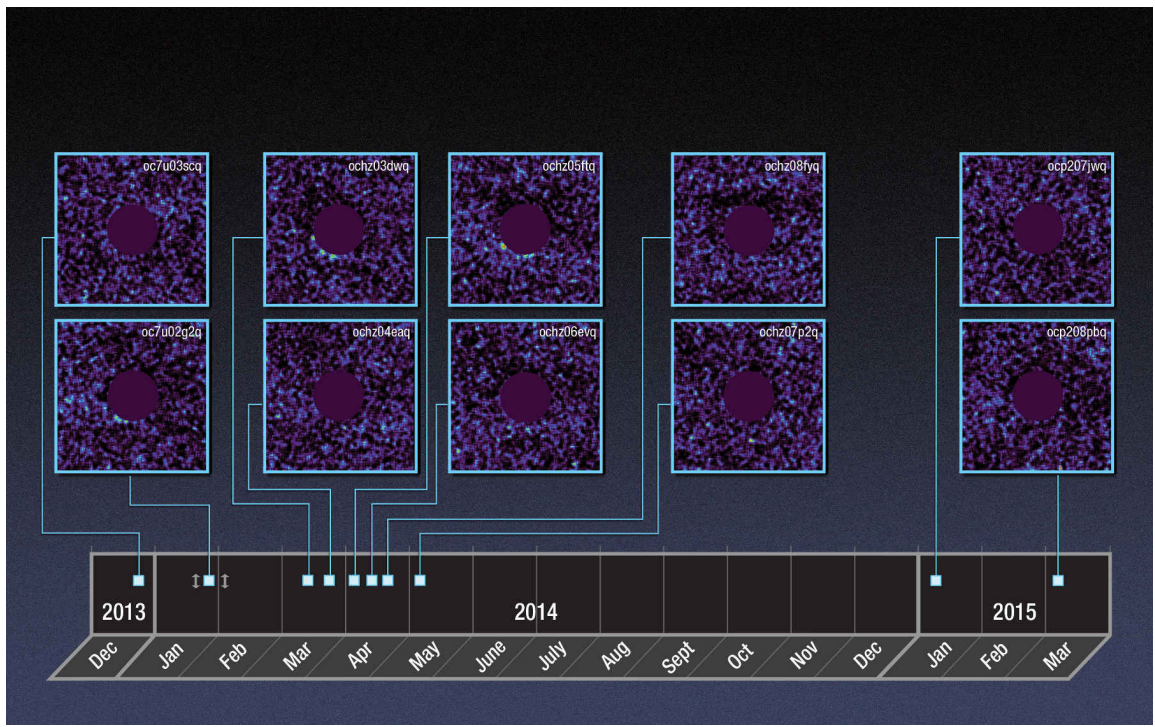

Figure 16

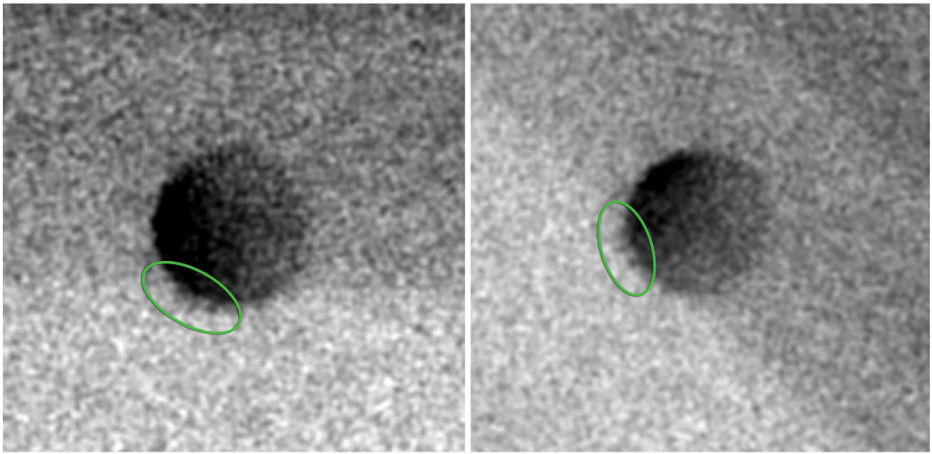

Figure 17

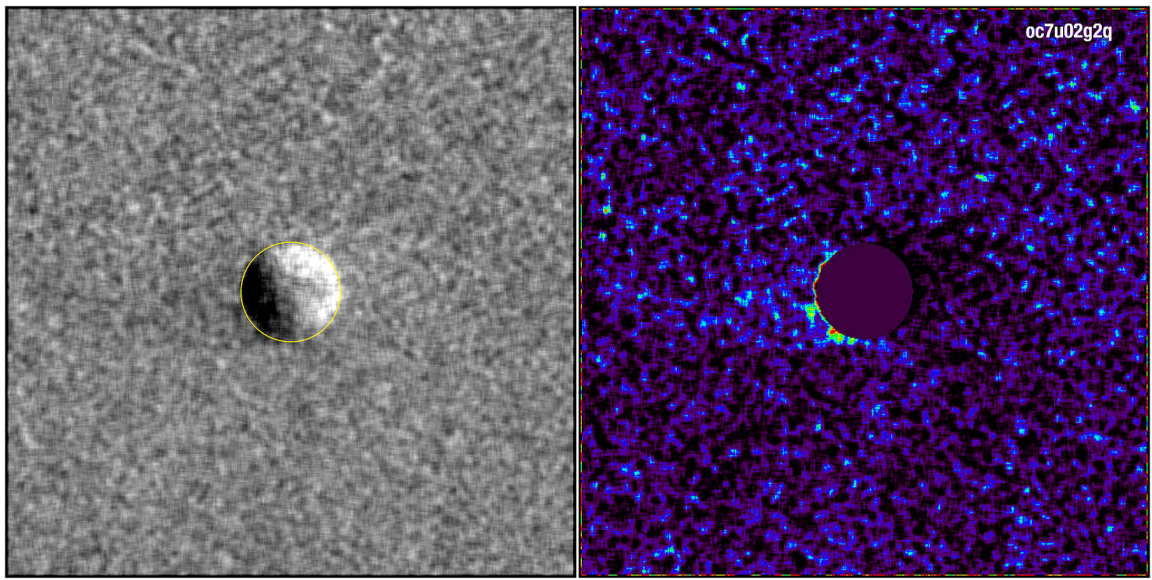

Figure 18

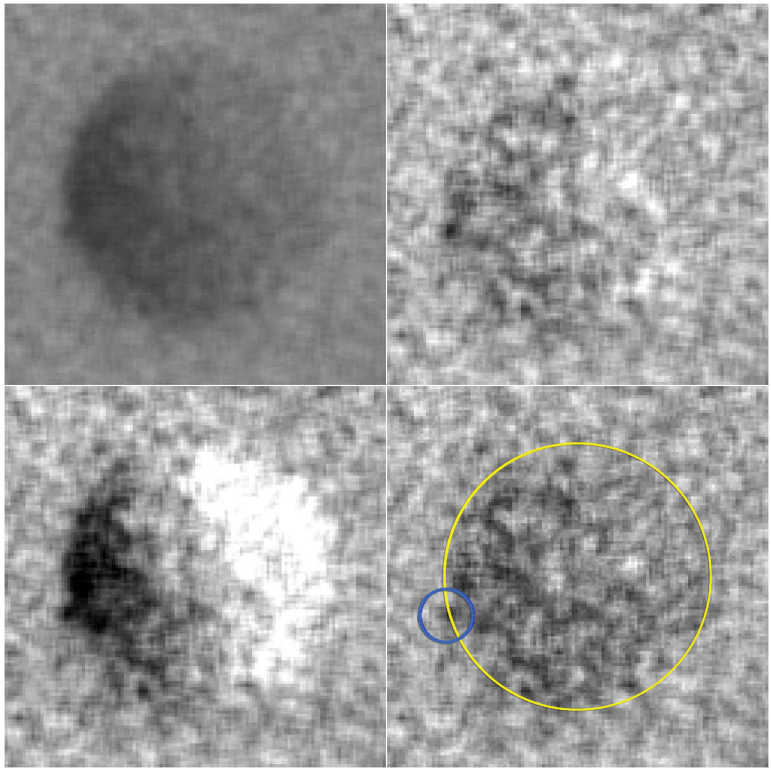

FIgure 19

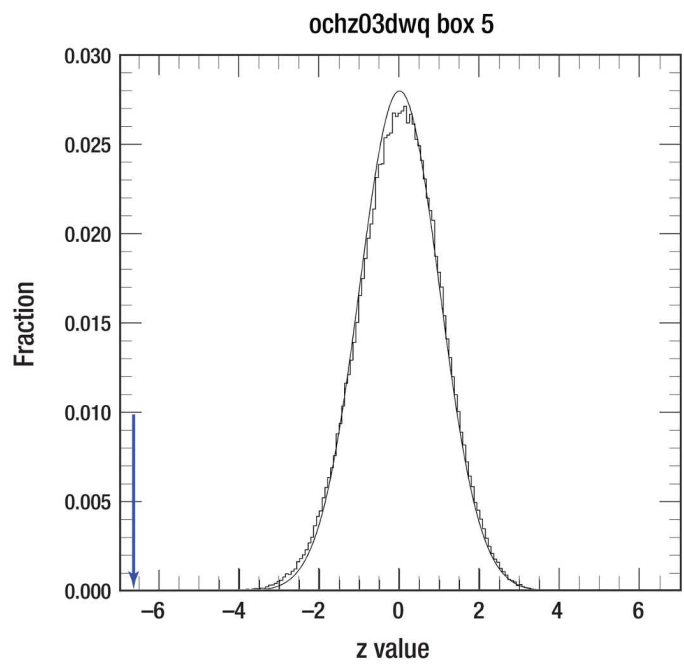

Figure 20

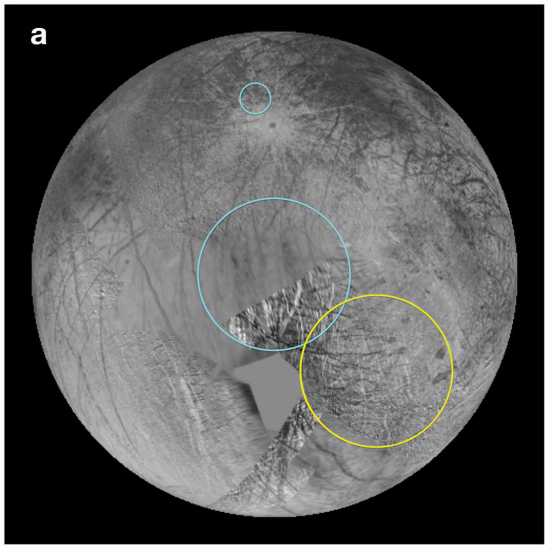
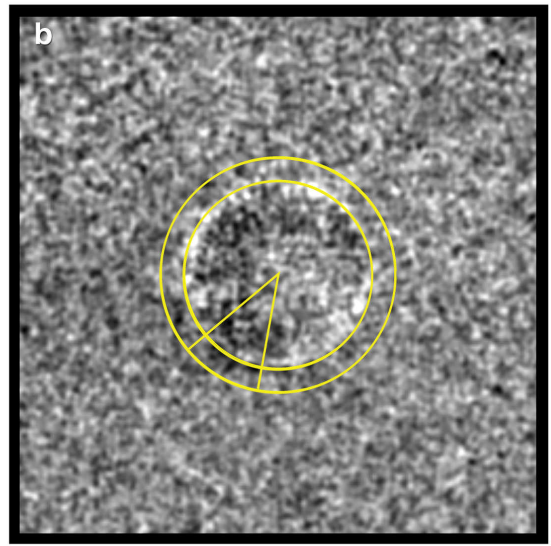

Figure 21

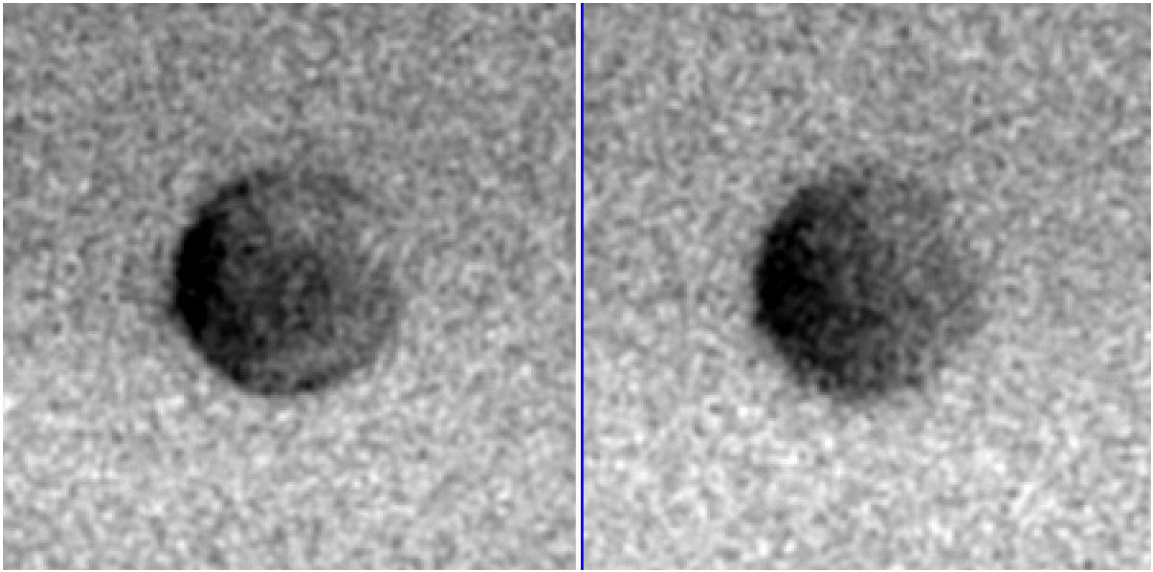

Figure 22

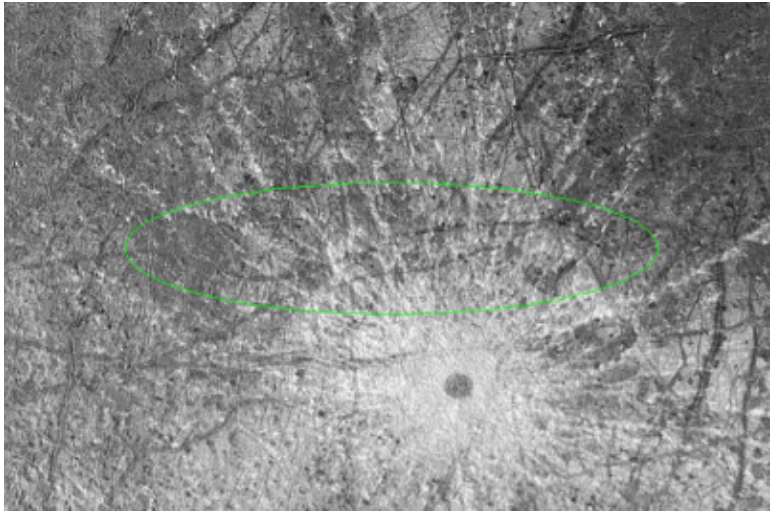

FIgure 23